\documentstyle[epsf,11pt]{article}
\newcommand{\bea}{\begin{eqnarray}}
\newcommand{\eea}{\end{eqnarray}}
\begin{document}

\newcommand{\lsim}
{{\;\raise0.3ex\hbox{$<$\kern-0.75em\raise-1.1ex\hbox{$\sim$}}\;}}
\newcommand{\gsim}
{{\;\raise0.3ex\hbox{$>$\kern-0.75em\raise-1.1ex\hbox{$\sim$}}\;}}

\begin{flushright}
{\large HIP-2001-63/TH\\
WU-ITP-2002.005} 
\end{flushright}
\vspace{0.1cm}

\begin{center}
{\LARGE\bf Sparticle spectrum and constraints in anomaly mediated
supersymmetry breaking models}
\\[15mm]
{\bf K. Huitu$^a$, J. Laamanen$^{a}$ and P.N. Pandita$^{a,b,c}$}\\[4mm]
$^a$Helsinki Institute of Physics\\
P.O. Box 64, FIN-00014 University of  Helsinki, Finland \\[4mm]
$^b$Institut f\"ur Theoretische Physik und Astrophysik\\
Universit\"at W\"urzburg, 97074 W\"urzburg, Germany\\[4mm]
$^c$Department of Physics, North Eastern Hill University
\\ Shillong 793 022,  India\footnote{Permanent address} \\[7mm]
\date{}
\end{center}

\begin{abstract}
We study in detail the particle spectrum in 
anomaly mediated supersymmetry breaking models in which supersymmetry
breaking terms are induced by the super-Weyl anomaly. We investigate
the minimal anomaly mediated supersymmetry breaking models, gaugino
assisted supersymmetry breaking models, as well as models with
additional residual nondecoupling $D$-term contributions due to an
extra U(1) gauge symmetry at a high energy scale. We derive sum rules
for the sparticle masses in these models which can help in
differentiating between them. We also  obtain the sparticle spectrum
numerically, and compare and contrast the results so obtained for the
different types of anomaly mediated supersymmetry breaking models.
\end{abstract}

\vskip 7mm

{~~~~PACS number(s): 12.60.Jv, 14.80.Ly}
\section{Introduction} Supersymmetry is at present the only framework in 
which the Higgs sector of the standard model (SM) is natural.  
It is, thus, a prominent candidate for physics beyond the SM.
Since in nature there are no supersymmetric particles with the same
mass as ordinary particles, supersymmetry must be a broken symmetry at
low energies.
The specific mechanism which breaks supersymmetry is important in
determining the sparticle masses and, hence, the experimental
signatures of supersymmetry.
At present there are several models of supersymmetry breaking. 
The model of supersymmetry breaking that has been studied most
extensively is the gravity mediated \cite{sugra} supersymmetry
breaking model.
In this class of models, supersymmetry is assumed to be broken in a
hidden sector by fields which interact with the SM particles and their
superpartners (the visible particles) only via gravitational
interactions.
Whereas this mechanism of supersymmetry breaking is simple and
appealing, it suffers from the supersymmetric flavor problem.
On the other hand, in a different class of models \cite{gmsb},
supersymmetry is broken in a hidden sector and transmitted to the
visible sector via SM gauge interactions of messenger particles.
This mechanism of supersymmetry breaking provides an appealing solution
to the supersymmetric flavor problem.
Both these types of supersymmetry breaking models have their
distinct experimental signatures.

The soft supersymmetry breaking terms in the above breaking mechanisms
have contributions originating from the super-Weyl anomaly via 
loop effects.
If gravity and gauge mediation are somehow suppressed, the anomaly mediated
contributions can dominate, as may happen,  {\it e.g.}, in brane models
\cite{amsb}. If this happens, then this  mechanism of supersymmetry
breaking is referred to as
anomaly mediated supersymmetry breaking (AMSB).
Anomaly mediation is a predictive framework for supersymmetry breaking
in which the breaking of scale invariance mediates between hidden and
visible sectors.

Since the soft supersymmetry breaking parameters are determined by
the breaking of the scale invariance, they can be written in terms of
the beta functions and anomalous dimensions in the form of relations
which hold at all energies.
In the minimal supersymmetric standard model (MSSM),
the pure anomaly mediated contributions to
the soft supersymmetry (SUSY) breaking parameters $M_\lambda$ (gaugino mass),
$m_{i}^2$ (soft scalar mass squared), and $A_y$ (the trilinear
supersymmetry breaking coupling, where $y$ refers to the Yukawa coupling) 
can be written as
\bea
M_\lambda &=& \frac{\beta_g}{g} m_{3/2},\label{gmass}\\
m_{i}^2 &=& -\frac 14 \left( \frac{\partial \gamma_i}{\partial
g}\beta_g + \frac{\partial \gamma_i}{\partial y}\beta_y\right)
m_{3/2}^2,\label{smass}\\
A_y &=& -\frac{\beta_y}{y} m_{3/2},\label{Amass}
\eea
where $m_{3/2}$ is the gravitino mass, 
$\beta$'s are the relevant $\beta$ functions, and $\gamma$'s are 
the anomalous dimensions of the chiral superfields.
An immediate consequence of these relations is that supersymmetry
breaking terms are completely insensitive to physics in the
ultraviolet.
The degrees of freedom that are excitable at a given energy, which
determine the anomalous dimensions and beta functions, thus
completely specify the soft supersymmetry breaking parameters at that
energy.
In this way the gaugino masses are proportional to their corresponding
gauge group $\beta$ functions with the lightest supersymmetric (SUSY)
particle being mainly a wino.
Analogously, the scalar masses and trilinear couplings are functions
of gauge and Yukawa-coupling $\beta$ functions.
However, it turns out that the pure scalar mass-squared anomaly
contribution for sleptons is negative~\cite{RS}.
There are a number of proposals for fixing this  problem of tachyonic
slepton masses \cite{PR,KSS,anomalyfix,JJ,chk}.
Additional contributions to the slepton masses
can arise in a number of ways, but some of the 
solutions will spoil the most attractive feature of the anomaly
mediated models, {\it i.e.}, the renormalization group (RG) invariance of
the soft terms and the consequent ultraviolet insensitivity of the
mass spectrum.
Nevertheless, there are various options to cure this problem without
reintroducing the supersymmetric flavor problem \cite{RS,PR}.
A simple phenomenologically  attractive way 
of parametrizing the
nonanomaly mediated contributions to the slepton masses, so as to
cure their tachyonic spectrum, is to add a common mass parameter $m_0$
to all the squared scalar masses \cite{GGW}, assuming that such an
addition does not reintroduce the supersymmetric flavor problem.
As noted above, such an 
addition of a nonanomaly mediated term destroys the attractive
feature of the RG invariance of soft masses.
However, the RG evolution of the resulting model, nevertheless,
inherits some of the simplicity of the pure anomaly mediated
relations.

There are several 
alternative ways to generate these extra contributions to the
soft squared masses.
In particular there are models of supersymmetry breaking mediated
through a small extra dimension, where SM matter multiplets and a
supersymmetry breaking hidden sector are confined to opposite
four-dimensional boundaries while gauge multiplets lie in the bulk.
In this scenario the soft gaugino mass terms are due to the 
anomaly mediated supersymmetry breaking.
On the other hand, 
scalar masses get contributions from both anomaly mediation and a tiny
hard breaking of supersymmetry by operators on the hidden sector
boundary.
These operators contribute to scalar masses at one loop and this
contribution is dominant, thereby making all squared scalar masses
positive.
The gaugino spectrum is unaltered, and the model resembles an anomaly
mediated supersymmetry breaking model with nonuniversal scalar masses
\cite{KK}.

Another class of models, where the problem with tachyonic slepton
masses is solved, is the models with additional residual
nondecoupling $D$-term contributions
due to extra U(1)'s at a high energy scale \cite{PR,KSS,JJ,chk}.
In these models one can preserve the property of having 
renormalization group invariant
soft terms, at least at the one-loop level \cite{chk,U1RGE}.
An interesting feature in this type of model is that unlike in the
minimal AMSB model, one can have a light stop  in the
spectrum~\cite{chk}. Furthermore, if the extra U(1) is anomaly-free, 
then it can be shown~\cite{JJ} that the ultraviolet insensitivity 
can be preserved to all orders.

In this paper  we consider the mass spectra, and the constraints
on this spectra,  of the anomaly mediated supersymmetric models.  
In Section 2 we study the sum rules for the scalar and gaugino
masses.
In Section 3 we obtain the focus points for the soft
scalar masses in the general anomaly mediated supersymmetric
models. This will help in determining whether large sparticle masses
are possible in these models without violating the
constraints of naturalness.
In Section 4  we present a detailed numerical study of the sparticle 
spectrum in different anomaly mediated supersymmetry breaking
models and compare and contrast them.
Section 5 is devoted to a discussion and summary of our results.

\section{Sum Rules}

The mass spectrum of superparticles in a particular supersymmetric 
model is determined in terms of soft supersymmetry breaking parameters.
These can be obtained at the weak scale by numerical solutions of the relevant
RG equations for a particular model with specific boundary conditions
at the high scale, usually taken to be the grand unified scale.
Since there are more supersymmetric particles than supersymmetry breaking
parameters, there are several relations between the sparticle 
masses, which can be written in terms of the sum rules~\cite{MR}.
These sum rules will,  in effect,  test the validity of a particular
supersymmetry breaking model. Thus by examining the relations
between the masses of sparticles, one may be able to distinguish between
different supersymmetric models. 
In this section we shall obtain various sum rules
involving sparticle masses for different anomaly mediated supersymmetry
breaking models.

\subsection{Scalar sector}

In the case of MSSM with gravity mediated supersymmetry breaking there
are seven physical scalar 
sparticle masses for the first two generations which
can be written in terms of four parameters (for  a given $\tan\beta =
v_2/v_1,  v_1$ and $v_2$ being the vacuum expectation values of the two
Higgs doublets of MSSM). This results in three sum rules for the
sparticle masses of the first two generations~\cite{MR}, which can be
used to test the various assumptions of MSSM with gravity mediated
supersymmetry breaking.

In anomaly mediated supersymmetry breaking models, the anomaly mediated
part of the soft masses is not running. Since for the first two
generations the Yukawa couplings can be neglected, we do not have any
contribution coming from the running of parameters to the masses of the
first two generations of squarks and sleptons. For the first two
generations, we can, therefore, write the physical masses of the squarks and
sleptons at any scale as (in the standard notation)
\bea
M_{\tilde u_L}^2 &=& c_Qm_0^2 +\left(\frac 12 -\frac 23
\sin^2\theta_W\right)M_Z^2\cos 2\beta + \left(-\frac{11}{50} g_1^4 -\frac 32
g_2^4 +8 g_3^4\right)\frac{m_{3/2}^2}{(16\pi^2)^2},\label{first}\\
M_{\tilde d_L}^2 &=& c_Qm_0^2 +\left(-\frac 12 +\frac 13
\sin^2\theta_W\right)M_Z^2\cos 2\beta + \left(-\frac{11}{50} g_1^4 -\frac 32
g_2^4 +8 g_3^4\right)\frac{m_{3/2}^2}{(16\pi^2)^2},\label{sec}\\
M_{\tilde u_R}^2 &=& c_um_0^2 +\frac 23
\sin^2\theta_W M_Z^2\cos 2\beta + \left(-\frac{88}{25} g_1^4
+8 g_3^4\right)\frac{m_{3/2}^2}{(16\pi^2)^2},\label{third}\\
M_{\tilde d_R}^2 &=& c_dm_0^2 -\frac 13
\sin^2\theta_W M_Z^2\cos 2\beta + \left(-\frac{22}{25} g_1^4
+8 g_3^4\right)\frac{m_{3/2}^2}{(16\pi^2)^2},\label{fourth}\\
M_{\tilde e_L}^2 &=& c_Lm_0^2 +\left(-\frac 12 +
\sin^2\theta_W\right)M_Z^2\cos 2\beta + \left(-\frac{99}{50} g_1^4 -\frac 32
g_2^4\right)\frac{m_{3/2}^2}{(16\pi^2)^2},\label{fifth}\\
M_{\tilde \nu}^2 &=& c_Lm_0^2 +\frac 12 M_Z^2\cos 2\beta +
\left(-\frac{99}{50} g_1^4 -\frac 32 g_2^4\right)
\frac{m_{3/2}^2}{(16\pi^2)^2},\label{sixth}\\
\label{last}
M_{\tilde e_R}^2 &=& c_em_0^2 -
\sin^2\theta_W M_Z^2\cos 2\beta + \left(-\frac{198}{25} g_1^4 \right)
\frac{m_{3/2}^2}{(16\pi^2)^2},
\eea
where we have parametrized the nonanomaly mediated contribution to the
masses via the parameter $m_0$, and we have assumed that this
contribution could be nonuniversal. Thus the parameters $c_Q, c_u,
c_d, c_L$ and $c_e$ could all be different from one another. In the
case of minimal anomaly-mediated supersymmetry breaking $c_Q =  c_u =
c_d = c_L = c_e = 1$. We have also included the $D$-term contribution
to the masses in Eqs. (\ref{first}) -- (\ref{last}). We can use these
equations to relate the masses of squarks and sleptons, via sum rules,
for different anomaly mediated supersymmetry breaking models. 

However, independently of the model,
Eqs.~(\ref{first}), (\ref{sec}) and 
(\ref{fifth}), (\ref{sixth}) lead to the sum rules
\bea
M_{\tilde d_L}^2-M_{\tilde u_L}^2 = -\cos 2\beta M_W^2, \label{indep1}\\
M_{\tilde e_L}^2-M_{\tilde \nu}^2 = -\cos 2\beta M_W^2,\label{indep2}
\eea
which relate the masses of squarks and sleptons living in the same SU(2)$_L$
doublet. 
We note that these sum rules are the same as in the gravity mediated
supersymmetry breaking models~\cite{MR}. 
These sum rules do not depend on the
assumption of universal soft breaking mass $m_0$, and depend only on the
$D$-term contribution to the squark and slepton masses. 
They are, thus, independent of the supersymmetry breaking model and test only 
the gauge structure of the effective  low energy supersymmetric model.
The other sum rules depend on the soft parameters originating from the
supersymmetry breaking mechanism, and are thus model dependent.
Depending on the coefficients multiplying $m_0$ in Eqs.~(\ref{first}) --
(\ref{last}), as specified by different  SUSY breaking models,
we have three additional sum rules.
In this section we consider in addition to the minimal anomaly
mediated, the gaugino assisted anomaly mediated SUSY breaking model
\cite{KK} and an AMSB model with additional U(1) and a light stop
\cite{chk}.

\subsubsection{Minimal anomaly mediated supersymmetry breaking model}

For the minimal model we have  $c_Q =  c_u = c_d = c_L = c_e = 1$. 
A third sum rule can then be obtained by taking a linear combination
of  Eqs.~(\ref{sec}) -- (\ref{sixth})  and  (\ref{last})
\begin{equation}
2(M_{\tilde u_R}^2 - M_{\tilde d_R}^2) +(M_{\tilde d_R}^2-M_{\tilde d_L}^2) +
(M_{\tilde e_L}^2 - M_{\tilde e_R}^2) =  \frac{10}{3} \sin ^2\theta_W
M_Z^2 \cos 2\beta ,\label{sum3}\\
\end{equation}
which is identical to the corresponding sum rule in the gravity mediated
models~\cite{MR}. This sum rule depends only 
on the  assumption of a universal
$m_0$, and is, therefore, a test of the universality of the soft scalar masses
in AMSB models as well. 

There are four remaining relations between the masses of the first two
generations of squarks and sleptons in Eqs.~(\ref{first}) -- (\ref{last}).
Two of these can be used to obtain expressions for the input parameters
$m_0$ and $m_{3/2}$ in terms of the squark and slepton masses.
Thus Eqs. (\ref{third}), (\ref{fourth}), (\ref{fifth}), and (\ref{last}) give
\bea
\label{mam0}
m_0^2 &=& M_{\tilde e_R}^2-\frac{44}{3}\tan^4\theta_W
(M_{\tilde e_L}^2 - M_{\tilde e_R}^2) -33\tan^4\theta_W 
(M_{\tilde u_R}^2 - M_{\tilde d_R}^2)
\nonumber\\&&
 +\left[-\frac{22}{3}\tan^4\theta_W+
\left(1+\frac{187}{3}\tan^4\theta_W\right)\sin^2\theta_W\right]
M_Z^2\cos 2\beta,\label{mamm0}\\
\frac 32 g_2^4{m_{3/2}^2\over{(16\pi^2)^2}} &=&
(M_{\tilde e_R}^2-M_{\tilde e_L}^2)-
\frac 94 (M_{\tilde u_R}^2 - M_{\tilde d_R}^2)-
\left(\frac 12-\frac{17}{4}\sin^2\theta_W\right)M_Z^2\cos 2\beta.
\label{mam32}
\eea

The remaining two independent equations can then be converted to two 
additional sum rules. The  first one is obtained from a combination
of Eqs. (\ref{third}),
(\ref{fourth}), and (\ref{last}):
\bea
&&(M_{\tilde e_L}^2 -M_{\tilde e_R}^2)+\frac 34\left(3-\frac{3}{11}
\cot^4\theta_W\right)(M_{\tilde u_R}^2-M_{\tilde d_R}^2)\nonumber\\
&&= \left[-\frac 12 + \left(\frac{17}{4}-\frac{9}{44}\cot^4\theta_W\right)
\sin^2\theta_W\right] M_Z^2\cos 2\beta .
\label{sum4}
\eea
It is well known that the left and right sleptons are almost
degenerate in the minimal AMSB model \cite{GGW}.
This can be easily verified explicitly from Eqs. (\ref{fifth})
and (\ref{last}),
\bea
(M_{\tilde e_L}^2 -M_{\tilde e_R}^2)\simeq -0.038 M_Z^2\cos 2\beta
-0.0078 M_2^2,
\label{sldiff}
\eea
where  we have used Eq.~(\ref{gmass}) for the gaugino mass parameters,
and the experimental value of $\sin\theta_W^2=0.2312$~\cite{PDG}.
Similarly one notices that in all the squark mass differences 
$M_{\tilde q_i}^2-M_{\tilde q_j}^2$ obtained from
Eqs.~(\ref{first}) -- (\ref{last}),  the contribution coming from the
$m_0$ part and the
strong coupling part cancel in the minimal AMSB model.  
Therefore such mass differences are small compared to the
masses themselves.

The mass difference between the right-handed squarks $\tilde u_R$ and 
$\tilde d_R$ {}from the sum rule (\ref{sum4}) is much larger than
Eq.~(\ref{sldiff}) for the slepton mass difference,
\bea
M_{\tilde u_R}^2-M_{\tilde d_R}^2\simeq 
-85.2 (M_{\tilde e_L}^2 -M_{\tilde e_R}^2)-3.44 M_Z^2\cos 2\beta
\simeq -0.2 M_Z^2\cos 2\beta+0.66 M_2^2,
\eea
where we have used Eq.~(\ref{sldiff}) to obtain
the last equality.
Although the coefficients of the two terms may not be small,
it should be
noticed that the part with $M_Z$ is constant and for heavy squarks
the mass difference between the $\tilde u_R$ and $\tilde d_R$ squarks
remains rather small.

The other independent sum rule can be obtained by combining Eqs.~
(\ref{first}), (\ref{sec}),
(\ref{fourth}), and  (\ref{last}) to give
\bea
&&(M_{\tilde e_L}^2 -M_{\tilde e_R}^2) +\left(\frac 94
-\frac{g_2^4}{2g_3^4}
\right) (M_{\tilde u_R}^2 -M_{\tilde d_R}^2)
-\frac{3}{16}\frac{g_2^4}{g_3^4} (M_{\tilde d_R}^2-M_{\tilde e_R}^2)
\nonumber\\
&&=\left[ -\frac 12+\left(\frac {17}{4}-\frac 58\frac{g_2^4}{g_3^4}\right)
\sin^2\theta_W\right]
M_Z^2\cos 2\beta.\label{sum5}
\eea
The sum rules (\ref{sum4}) and (\ref{sum5}) are unique
to the minimal anomaly mediated supersymmetry breaking model.

\subsubsection{Gaugino assisted AMSB model}

In the gaugino assisted anomaly mediated  model, it is assumed that
the gauge and gaugino fields reside in the bulk of the extra
dimension \cite{KK}.
The hidden sector is not supposed to contain singlets in which case
the anomaly mediated contribution is the dominant contribution 
to the supersymmetry breaking.
As a result of the gaugino wave function renormalization and
consequent rescaling of the fields, the scalar soft masses receive
contributions proportional to the eigenvalues of the
quadratic Casimir operators of the relevant
gauge group.
In this case the coefficients of
$m_0$ in Eqs. (\ref{first}) -- (\ref{last}) are \cite{KK}
\begin{equation}
c_Q=21/10,\; c_u=8/5,\; c_d=7/5,\; c_L=9/10,\; c_e=3/5.
\label{gcval}
\end{equation}
In the gaugino assisted AMSB model, the parameters $m_0$ and $m_{3/2}$ 
can be written in the form
\bea
m_0^2 &=& \frac 53 (M_{\tilde e_R}^2+\sin^2\theta_W M_Z^2\cos 2\beta)
+\left(\frac {9}{55}\cot^4\theta_W-3\right)^{-1}\left[
5M_{\tilde e_R}^2 \right.
\nonumber\\&&
\left.-4 (M_{\tilde e_L}^2 -M_{\tilde e_R}^2)-
9(M_{\tilde u_R}^2 -M_{\tilde d_R}^2)+
(-2+22\sin^2\theta_W) M_Z^2\cos 2\beta\right],\label{gm0}
\eea
\bea
6\left(g_2^4-\frac{33}{5}g_1^4\right){m_{3/2}^2\over{(16\pi^2)^2}}&=&
5M_{\tilde e_R}^2-4(M_{\tilde e_L}^2 -M_{\tilde e_R}^2)-
9(M_{\tilde u_R}^2 -M_{\tilde d_R}^2)\nonumber\\
&&+(-2+22\sin^2\theta_W) M_Z^2\cos2\beta.
\eea

\begin{figure}[t]
\leavevmode
\begin{center}
\mbox{\epsfxsize=8.5truecm\epsfysize=8.5truecm\epsffile{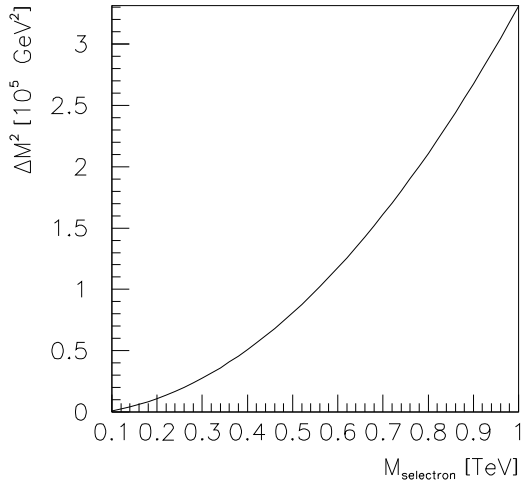}}
\end{center}
\caption{\label{figsum4} The difference 
$\Delta M^2\equiv M_{\tilde u_R}^2 -M_{\tilde d_R}^2$ as a function of the
right-handed selectron mass $M_{\tilde e_R}$.
We have plotted the curve for $\tan\beta=$10, but
the curves for  $\tan\beta>3$ are all almost
identical.
}
\end{figure}
In the previous section it was stated that the sum rule (\ref{sum3})
is a test of universality of the mass parameter $m_0$, since in
the universal case the dependence on $m_0$ cancels.
Although the extra contributions to the scalar masses are not
universal in the gaugino assisted AMSB model, the dependence on
the $m_0$ accidentally cancels, and the sum rule (\ref{sum3})
is valid also in the gaugino assisted AMSB model.
Thus even if the physical masses satisfy Eq. (\ref{sum3}), one needs
further confirmation of the universality of the extra contributions.
In the present case, a fourth sum rule, corresponding to Eq. (\ref{sum4}),
can be obtained as  
\bea
3(M_{\tilde u_R}^2 -M_{\tilde d_R}^2)-M_{\tilde e_R}^2
=4\sin^2\theta_W M_Z^2\cos 2\beta.\label{gsum4}
\eea
The simple form of this sum rule follows from the fact that the
dependence on $m_0$ vanishes also in this case.
Since the experimental lower bound on the selectron and smuon masses
is higher than the $Z$-boson mass, one can deduce from Eq. (\ref{gsum4})
that the right-handed up-squark is heavier than the right-handed
down-squark.
This can also be seen from Fig.~\ref{figsum4}, where
the difference $M_{\tilde u_R}^2 -M_{\tilde d_R}^2$ is plotted as a 
function of the selectron mass $M_{\tilde e_R}$.
The curves for  $\tan\beta>3$ are all practically 
identical to the curve shown
here, since the dependence on $\tan\beta$ is rather weak.
This differs from the minimal AMSB model, where 
$M_{\tilde u_R}^2 -M_{\tilde d_R}^2$ depends on $M_2^2$
instead of a scalar mass.

The last sum rule, corresponding to the sum rule (\ref{sum5})
of the minimal anomaly mediated supersymmetry breaking model, 
can, in this case, be written as
\bea
&&5\left( 1+\frac r3\right)M_{\tilde e_R}^2-
4(M_{\tilde e_L}^2 -M_{\tilde e_R}^2)
-(9+2r)(M_{\tilde u_R}^2 -M_{\tilde d_R}^2)-\frac{3r}{4}
(M_{\tilde d_R}^2 -M_{\tilde e_R}^2)\nonumber\\
&&=\left[ 2-\left(\frac{25 r}{6}
+22\right)\sin^2\theta_W\right]M_Z^2\cos 2\beta,
\eea
where $r$ is given by $r\equiv (g_2^4-\frac{33}{5} g_1^4)/(\frac{11}{5}g_1^4
-g_3^4)$.

\subsubsection{AMSB model with additional U(1) and a light stop}

An interesting possible solution to the tachyonic slepton mass problem 
is to modify the gauge group of the theory from that of
the MSSM gauge group by an additional U(1) 
factor group \cite{PR,KSS,JJ,chk}.
In the case of MSSM, it is well known that the $D$-term contributions 
to the scalar masses are of opposite sign for left- and right-handed 
particles.  However, if the left- and right-handed sleptons both had
positive charge under this additional  U(1) gauge group, then we could 
have positive contributions to the slepton masses from $D$ terms, 
and possibly solve the tachyonic slepton problem.
Furthermore, by selecting the charges suitably, one may also
have renormalization group invariant soft terms, at least at the 
one-loop level \cite{chk,U1RGE}. Sum rules for a model with 
Fayet-Iliopoulos $D$ terms due to extra U(1),
which are independent of $M_W$ and $M_Z$, were derived in~\cite{JJ}.

In Ref.~\cite{chk}, models with an extra U(1) and RG-invariant sum 
of the squares of
soft SUSY breaking scalar masses were considered.
In that paper two models were explicitly constructed, one
with a spectrum similar to the minimal AMSB models and another model 
with a light stop.
It turns out that the U(1) charge assignment for getting a light stop
is quite constrained.
One particular set of the U(1) charges that leads to a light stop
is the following:
\bea
c_Q=3,\; c_u=-1,\; c_d=-1,\; c_L=1,\; c_e=1.
\label{ucval}
\eea
Here we will consider the sum rules for this particular light stop model.
Since $c_u=c_d$, as well as $c_L=c_e$, the formulas (\ref{mamm0}) and 
(\ref{mam32}) of the minimal anomaly mediated supersymmetry
breaking model are valid in this model as well.
Note that the negative soft terms for right-handed squarks are the
reason behind the light right-handed stop.

On the other hand, the sum rule (\ref{sum3}) is no longer  valid.
When we take a similar combination as in Eq. (\ref{sum3})
of the soft mass parameters,
and use the expression for $m_0$, we find
\bea
&&\left(1-\frac{176}{3}\tan^4\theta_W\right)
(M_{\tilde e_L}^2 - M_{\tilde e_R}^2)+\left(2-132
\tan^4\theta_W\right)(M_{\tilde u_R}^2 - M_{\tilde d_R}^2)
+(M_{\tilde d_R}^2-M_{\tilde d_L}^2)\nonumber\\
&&+4M_{\tilde e_R}^2=-\frac 23\left[-44\tan^4\theta_W+
\left(1 + 374\tan^4\theta_W\right)\sin^2\theta_W\right]M_Z^2\cos 2\beta.
\label{u11}
\eea
The sum rule (\ref{sum4}) of the minimal AMSB model is valid in this 
model as well.
This is easily verified by taking an appropriate combination of Eqs.
(\ref{third}), (\ref{fourth}), and (\ref{last}), and noticing that
in both models $c_u = c_d$, $c_e = 1$, and that 
the parameter $m_0$ is determined via Eq.~(\ref{mam0}) 
in both these models.
The arguments given in section 2.1.1 for the degenerate slepton masses
and small mass difference $M_{\tilde u_R}^2 -M_{\tilde d_R}^2$
in minimal AMSB are valid in this model as well.
{}From the sum rule (\ref{u11}) one can write approximately
\bea
M_{\tilde d_R}^2-M_{\tilde d_L}^2\simeq 4.3 
(M_{\tilde e_L}^2 - M_{\tilde e_R}^2) +9.9
(M_{\tilde u_R}^2 - M_{\tilde d_R}^2) -4M_{\tilde e_R}^2
-2.72 M_Z^2\cos 2\beta.
\label{dlrdiff}
\eea
One can easily check that for experimentally allowed parameter values,
the right-hand side of Eq. (\ref{dlrdiff}) is always negative and not
small, and it decreases 
with the increasing selectron mass.

To complete the derivation of independent sum rules of this case we
give the sum rule corresponding to Eq. (\ref{sum5}):
\bea
&&(M_{\tilde d_R}^2 - M_{\tilde e_R}^2)+\frac 83
(M_{\tilde u_R}^2 - M_{\tilde d_R}^2)-\left(\frac 23
\frac{g_3^4}{g_2^4}+\frac{11}{3}\tan^4\theta_W\right)
\left[8(M_{\tilde e_L}^2 - M_{\tilde e_R}^2) +18
(M_{\tilde u_R}^2 - M_{\tilde d_R}^2)\right]\nonumber\\
&&+2M_{\tilde e_R}^2=\left[\left(\frac 23
\frac{g_3^4}{g_2^4}+\frac{11}{3}\tan^4\theta_W\right)
(4-34\sin^2\theta_W)+\frac 43\sin^2\theta_W\right]M_Z^2\cos 2\beta.
\label{u12}
\eea

\subsubsection{Third generation of scalars}

So far we have considered the first two generations of squarks and 
sleptons for which the corresponding Yukawa couplings and their runnings
can be neglected. For the third family squarks and sleptons the
sum rules are complicated because of the large 
third generation Yukawa couplings.
For small values of $\tan\beta$, we can neglect the effects of
the bottom-quark Yukawa coupling, and hence the effects of mixing in the
bottom-squark mass matrix. Thus in this limit $\tilde b_L$ and 
$\tilde b_R$ are still the mass eigenstates, and $\tilde b_R$
is degenerate with $\tilde d_R$ to a good approximation.
However, since the evolution of $M_{\tilde b_L}$ is controlled by the 
top-quark Yukawa coupling, the situation with respect to hierarchy of 
the sbottom masses and the squark masses of the first two generations
in anomaly mediated supersymmetry breaking models can be predicted 
only when we have a detailed knowledge of input parameters.
In the stop sector, it is easy to see that  
the sum rules one obtains are independent of the model of supersymmetry 
breaking, and are, thus, identical to those obtained in MSSM 
with gravity mediated supersymmetry breaking~\cite{MR}.
Similar observations can be made with respect to the sum rules
in the stau sector.  Furthermore, when $\tan\beta$ is large, 
the mixing in the third generation
squark and slepton mass matrices becomes
important, and the situation becomes complicated.
Thus, as far as the third generation squarks and sleptons
are concerned, one does not obtain any information which could 
distinguish the anomaly mediated supersymmetry breaking models
from the MSSM with gravity mediated supersymmetry breaking.

\subsection{Gaugino sector}

In all the models discussed in this work, the gaugino sector remains 
the same as in the minimal AMSB model, for which the
mass difference between the lightest chargino and the neutralino is
small.
The close proximity of the lightest neutralino and chargino masses is
a direct consequence of  Eq. (\ref{gmass}), which gives for
the ratios of the gaugino mass parameters 
$|M_1|:|M_2|:|M_3|\simeq 2.8:1:7.1$ after taking into account the next to
leading order radiative corrections
and the weak scale threshold corrections \cite{GGW}.
Thus the winos are the lightest neutralinos and charginos, and
one would expect that the lightest chargino is only slightly heavier
than the lightest neutralino in all the models considered in Section 2.1.

It is not feasible to obtain mass sum rules for the neutralino states,
since the physical neutralino mass matrix is a
$4\times4$ matrix.
However, from the trace of neutralino and chargino mass matrices,
one obtains a sum rule,  which does not contain the Higgs mixing
parameter $\mu$, but  which is present in the mass matrices
[here the gluino mass $m_{\tilde g}=M_3 (t_{\tilde g})$],
\bea
2(M_{\tilde \chi^\pm_1}^2+M_{\tilde \chi^\pm_2}^2)-(M_{\tilde \chi^0_1}^2+
M_{\tilde \chi^0_2}^2+M_{\tilde \chi^0_3}^2+M_{\tilde
\chi^0_4}^2)
\nonumber\\
=\frac 19
\left[\frac{g_2^4}{g_3^4}-\left(\frac{33}{5}\right)^2
\frac{g_1^4}{g_3^4}\right]m_{\tilde g}^2 +4M_W^2-2M_Z^2.
\label{gauginosum}
\eea
This average mass squared difference of the charginos and neutralinos
differs from the corresponding supergravity (SUGRA) result
by the factor $(33/5)^2$ in front of
$g_1^4$ within the square brackets, 
and by factor (1/9) in front of the square bracket.
We note that the 
ratio of the gluino mass to the other gaugino masses  is different
in the AMSB models and in the MSSM with gravity mediated
supersymmetry breaking,
\bea
\left.\frac{|m_{\tilde g}|}{|M_2|}\right|_{\rm AMSB}
=3 \left.\frac{|m_{\tilde g}|}{|M_2|}\right|_{\rm MSSM}.
\eea
\begin{figure}[t]
\leavevmode
\begin{center}
\mbox{\epsfxsize=8.5truecm\epsfysize=8.5truecm\epsffile{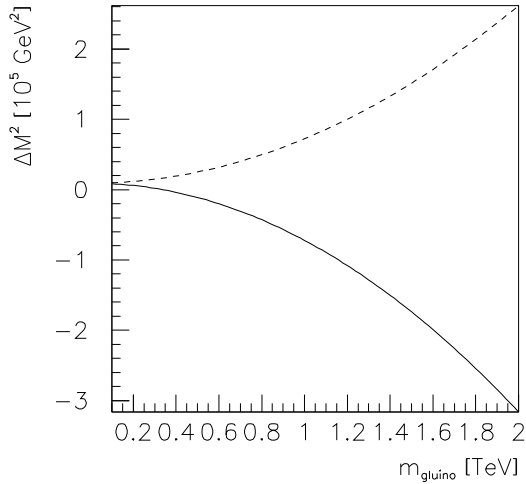}}
\end{center}
\caption{\label{gauginos} The average mass difference 
$\Delta M^2\equiv 2\sum M_{\chi^\pm_i}^2-\sum M_{\chi^0_i}^2$ 
in the AMSB models (solid line), 
and in the minimal SUGRA (dashed line) as a function of the gluino mass 
$m_{\tilde g}$.}
\end{figure}
We have plotted in Fig.~\ref{gauginos} the sum rule (\ref{gauginosum})
both in the AMSB models and the MSSM.
The average mass difference in the AMSB models is first positive, 
but then quickly turns negative (solid line), while in the minimal 
SUGRA model it is always positive (dashed line).
Thus this sum rule could be  one of the signatures of the AMSB 
type of models.

\section{Focus points}

One of the main motivations for low scale supersymmetry is that 
it can  stabilize the large hierarchy between the weak scale and the
unification or Planck scale. This can be realized in a straightforward 
way if the masses of the supersymmetric partners of the standard model 
particles are of the order of weak scale. On the other hand the existence 
of superpartners with masses of the order of weak scale
is difficult to reconcile with limits on flavor changing processes, 
unless one assumes an accurate degeneracy of squarks and sleptons.
Even with such a degeneracy, the supersymmetric contributions to
$CP$ violation become uncomfortably large, unless squarks and sleptons 
are heavy. Thus one is forced to examine supersymmetric 
models with large sparticle masses, thereby coming into conflict
with the basic idea of introducing weak scale supersymmetry.
Recently, it has been pointed out~\cite{FMM} that one can have 
large squark and slepton masses (above $1$ TeV) without losing
the naturalness of the underlying supersymmetric theory.
This is achieved by exploiting the existence of focus points
in the renormalization group evolution of the soft masses, which
make the weak scale insensitive to the variations in the unknown
supersymmetry breaking parameters at the high scale. 
In this approach the
squark and slepton masses can be large compared to the weak scale, 
though gaugino and higgsino masses can be generically lighter.
It is, therefore, important to investigate whether the anomaly
mediated supersymmetry breaking models considered in this paper
exhibit the focus point behavior so as to admit heavy squark and 
slepton masses, without violating the principle of naturalness.

In the minimal AMSB model, it has been  shown that
there is a focus point 
near the weak scale for the soft supersymmetry breaking
Higgs mass parameter $m_{H_u}^2$ if $\tan\beta$
is not very large~\cite{FM}.
The other mass parameters in this model 
do not have focus points near the weak
scale.  Since in the more general models studied in this work the
soft supersymmetry breaking mass parameters are not universal, 
it is of considerable interest to find out
if there is focus point behavior in these models. In this section we shall
consider this question and investigate if these general
models exhibit a desirable focus point behavior.

We shall here consider the case of 
$\tan\beta$=10, for which all couplings other than the 
top Yukawa coupling can be neglected. 
Following \cite{FM} we denote the supersymmetry
breaking bilinear up-type Higgs mass parameter
$m_{H_u}^2\equiv m_{H_u}^2|_{AM}+
\delta m_{H_u}^2$, where $m_{H_u}^2|_{AM}$ is the pure anomaly mediated
value, which does not run, and a similar decomposition
for the other scalar mass parameters.
Then the coupled  renormalization group equations 
for the relevant mass parameters are given by 
\bea
\frac{d}{dt}\left(\begin{array}{c} 
\delta m_{H_u}^2\\\delta m_{U_3}^2\\\delta m_{Q_3}^2
\end{array}\right) = \frac{Y_t^2}{8\pi^2}
\left(\begin{array}{ccc}
3&3&3\\2&2&2\\1&1&1 
\end{array}\right)
\left(\begin{array}{c} 
\delta m_{H_u}^2\\\delta m_{U_3}^2\\\delta m_{Q_3}^2
\end{array}\right).
\eea
As was done for sfermions in Eqs.~(\ref{first}) -- (\ref{last}), 
one can define the coefficients $c_{H_u}$ and
$c_{H_d}$ which parametrize the nonuniversality for the 
mass parameters of the  two Higgs doublets.
Here we need only $c_{H_u}$, which has a value $c_{H_u}=1$
in the minimal AMSB model, a value $c_{H_u}=9/10$
in the gaugino assisted AMSB model, 
and a value $c_{H_u}=-2$ in the model with an extra U(1) and a light stop.
For a set of general initial conditions, $m_0^2 (c_{H_u},c_{U_3},c_{Q_3})$,
the solution can be written as 
\bea
\left(\begin{array}{c} 
\delta m_{H_u}^2\\\delta m_{U_3}^2\\\delta m_{Q_3}^2
\end{array}\right) &=& \frac{(c_{H_u}+c_{U_3}+c_{Q_3})}{6}
m_0^2\exp\left[6\int_0^t dt'\frac{Y_t^2}{8\pi^2}\right]
\left(\begin{array}{c}3\\2\\1\end{array}\right)\nonumber \\
&&+\frac{m_0^2}{6} \left(\begin{array}{c} 
3(c_{H_u}-c_{U_3}-c_{Q_3})\\2(-c_{H_u}+2c_{U_3}-c_{Q_3})\\
(-c_{H_u}-c_{U_3}+5c_{Q_3})
\end{array}\right).
\eea
It is easy to verify that $m_{H_u}$, $m_{U_3}$, and $m_{Q_3}$ can all
have focus points only if $c_{H_u}=c_{U_3}=c_{Q_3}=0$.
Thus this case is not of interest here.
One can also find conditions that need to be satisfied if
two of the mass parameters have focus points.
Below we list the two mass parameters that have
simultaneous focus points,  together 
with the relevant conditions that need to be satisfied (here 
$\alpha_t = \exp [6\int^t_0 \,dt'\,Y_t^2/(8\pi^2)]$):
\bea
&m_{H_u},\; m_{U_3}:&\; \; c_{H_u}=\frac 32
c_{U_3}=\frac{3(1-\alpha_t)}{5\alpha_t+1}c_{Q_3}, \nonumber \\
&m_{H_u},\; m_{Q_3}:&\; \; c_{H_u}=3
c_{Q_3}=\frac{3(1-\alpha_t)}{2(2\alpha_t+1)}c_{U_3}, \nonumber \\
&m_{U_3},\; m_{Q_3}:&\; \; c_{U_3}=2
c_{Q_3}=\frac{2(1-\alpha_t)}{3(\alpha_t+1)}c_{H_u}. \nonumber 
\eea
Obviously these conditions are not realized for the models that  we
have considered in this paper.
Finally, the simple case of a focus point for $m_{H_u}$ is found if
$c_{H_u}<c_{U_3}+c_{Q_3}$, for $m_{U_3}$ if
$2c_{U_3}<c_{H_u}+c_{Q_3}$, and for $m_{Q_3}$ if
$5c_{Q_3}<c_{H_u}+c_{U_3}$.
In these inequalities it is assumed that $c_{H_u}+c_{U_3}+c_{Q_3}>0$.
If this sum is negative, then these  inequalities  should be
reversed. We see that for both 
the minimal and gaugino assisted AMSB models we
have a focus point for $m_{H_u}$.
The case of the extra U(1) model is qualitatively different, since
$c_{H_u}+c_{U_3}+c_{Q_3}=0$ by construction in this model, and
thus it is not relevant
to state at which scale the boundary conditions are given.

\begin{figure}[t]
\leavevmode
\begin{center}
\mbox{\epsfxsize=9.truecm\epsfysize=9.truecm\epsffile{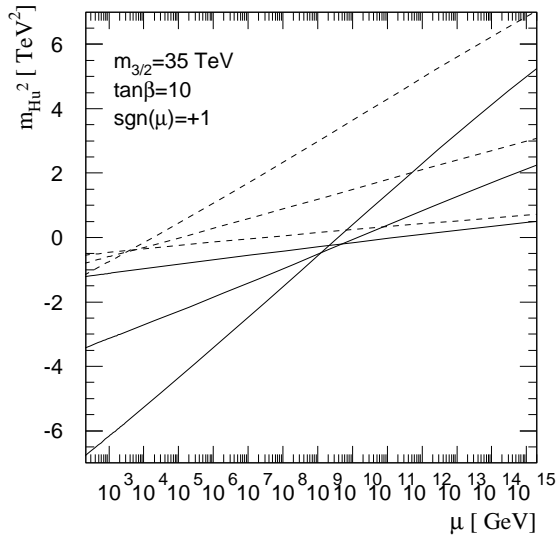}}
\end{center}
\caption{\label{focus} 
Focus points for the minimal (dashed line) and gaugino assisted 
(solid line) AMSB models.}
\end{figure}
In the gaugino assisted AMSB model the expression
$\exp\left[6\int_0^t dt'\frac{Y_t^2}{8\pi^2}\right]=14/23$ at the
focus point.
In Fig. \ref{focus} we  depict the running of $m_{H_u}^2$
for the minimal and for the gaugino assisted AMSB models.
Although focus points exist for both these models, 
only the focus point for the  minimal AMSB seems to be physically
interesting.

\section{Numerical results}

So far we have obtained the predictions for the anomaly mediated 
supersymmetry breaking models in terms of sum rules, which can help
in distinguishing between different models. In this section we shall
numerically obtain the predictions for the sparticle spectra of 
these models. For the numerical evaluation of the spectra 
we have used the program \verb|SOFTSUSY|~\cite{softsusy}. 
This program uses complete three family mass matrices
and Yukawa coupling evolution.
Fermion masses and gauge couplings ($\alpha , \alpha_s$) are evolved 
to $M_Z$ with the three loop QCD and one loop QED equations. The full MSSM 
spectrum is taken into account 
and decoupling threshold effects are taken into account 
to leading logarithmic order as well as the 
finite corrections at the scale $M_Z$. The scalar masses are
evolved via one-loop RG equations, and all other 
functions are evolved via two-loop equations.
Parameters are determined iteratively.
For the boundary conditions at the GUT scale, we have used
the values given by Eqs.
(\ref{gmass}), (\ref{Amass}), and $m_i^2 + c_i m_0^2$, where
$m_i^2$ is given by Eq. (\ref{smass}),
and $c_i $'s are model dependent numerical factors, which are given, 
{\it e.g.}, by Eqs. (\ref{gcval}) and (\ref{ucval}).

\begin{figure}[t]
\leavevmode
\begin{center}
\mbox{\epsfxsize=10.truecm\epsfysize=10.truecm\epsffile{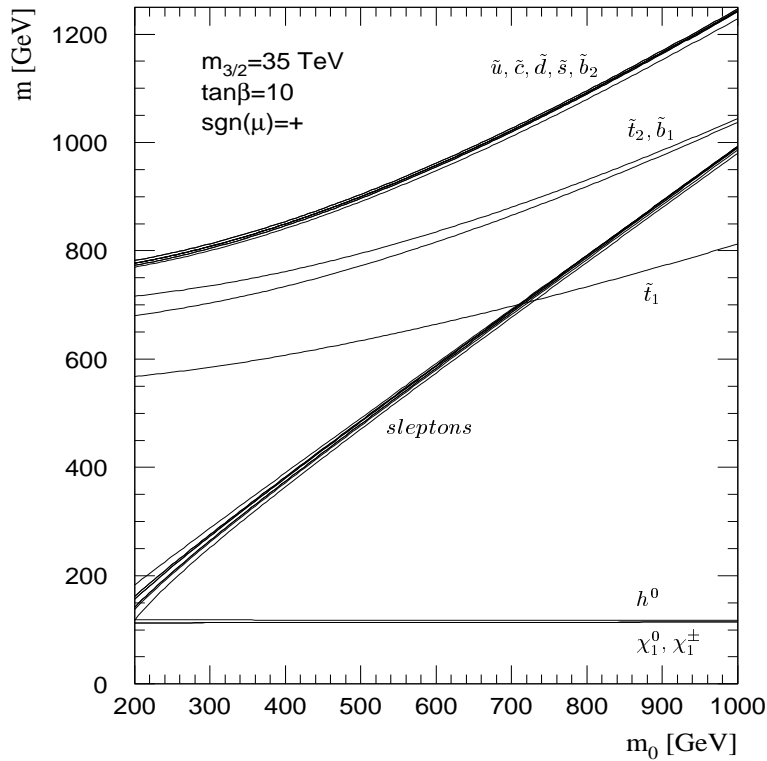}}
\end{center}
\caption{\label{spec1} The spectrum of the minimal AMSB model.}
\end{figure}
\begin{figure}[t]
\leavevmode
\begin{center}
\mbox{\epsfxsize=10.truecm\epsfysize=10.truecm\epsffile{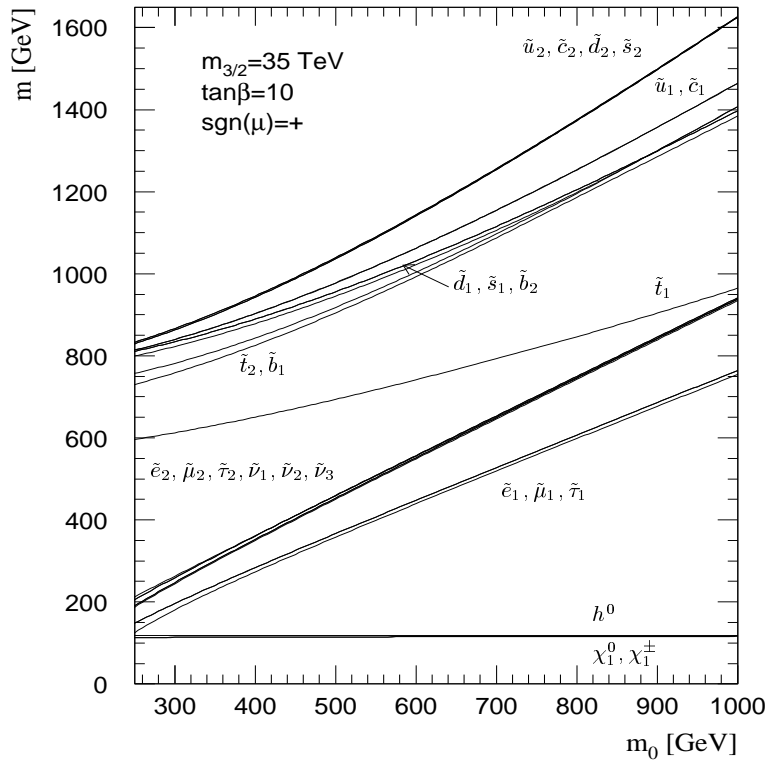}}
\end{center}
\caption{\label{spec2} The spectrum of the gaugino assisted AMSB model.}
\end{figure}
\begin{figure}[t]
\leavevmode
\begin{center}
\mbox{\epsfxsize=10.truecm\epsfysize=10.truecm\epsffile{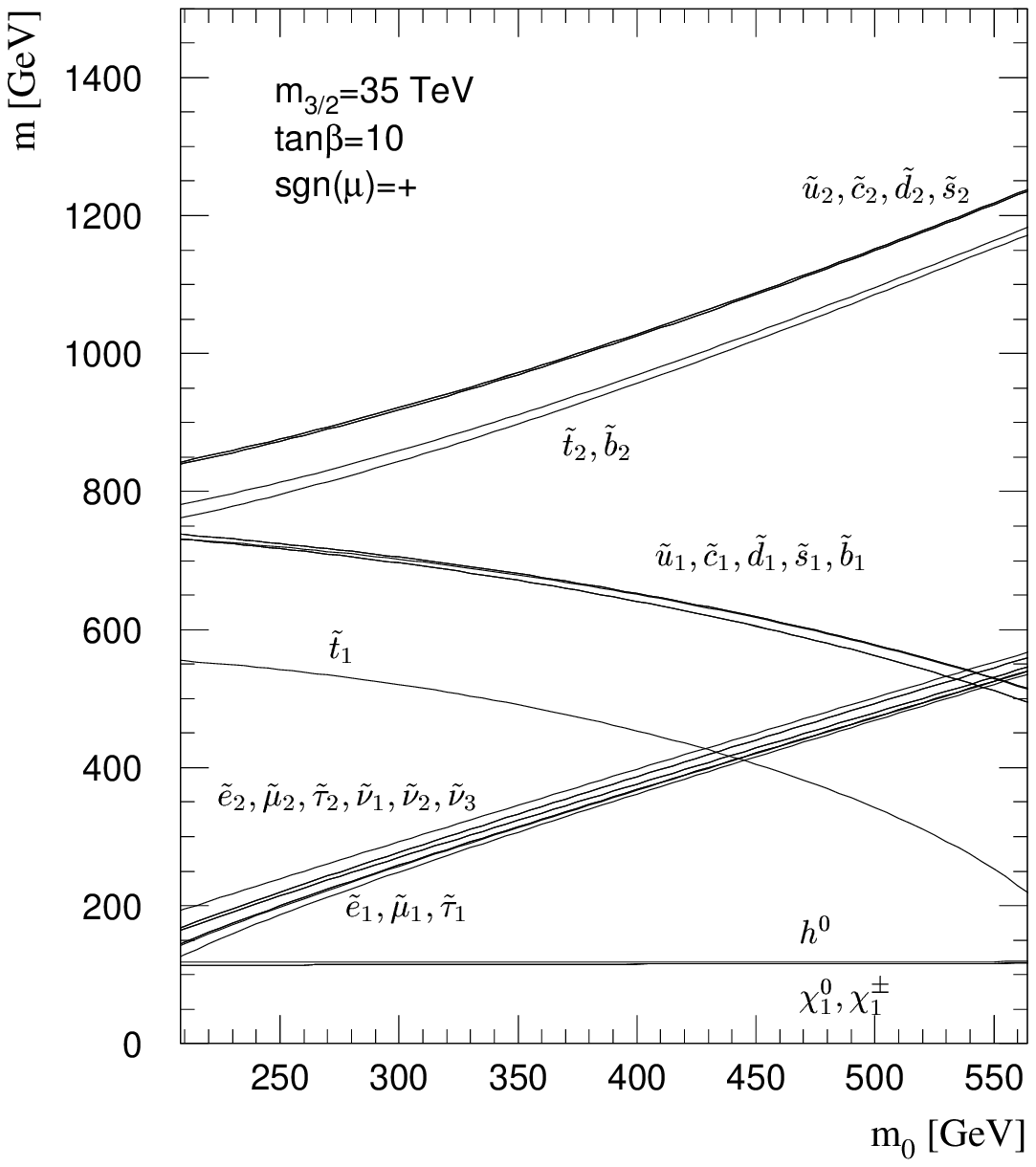}}
\end{center}
\caption{\label{spec3} The spectrum of the U(1) model.}
\end{figure}
In Figs. \ref{spec1} -- \ref{spec3} we have
plotted the spectra for the three different AMSB models.
In all three 
examples we have taken $\tan\beta=10$ and $m_{3/2}=35$ TeV.
One can easily see the distinguishing features of different models
already discussed  in terms of the sum rules.
In the minimal AMSB model, Fig.~\ref{spec1},
the squarks of the first two generations are
very closely degenerate in mass. The same is true for the sleptons
in this model. The third generation, especially the lighter stop, 
is considerably lighter than the  squarks of the first two generations.  
For $m_{3/2}=35 $ TeV we see that for a value of
$m_0\sim 730$ GeV, the lightest stop becomes lighter than sleptons.

For the gaugino assisted AMSB model, Fig.~\ref{spec2},
the first two generations of 
sleptons are not degenerate anymore.
Recalling that $\tilde u_1\; (\tilde c_1)$, $\tilde d_1\; 
(\tilde s_1)$, and $\tilde e_1\; (\tilde \mu_1)$ correspond
almost exactly to the right-handed squarks and sleptons,
we note that the sum rule Eq. (\ref{gsum4}) is clearly satisfied.
On the other hand, the left-handed squarks, corresponding to
$\tilde u_2$ and $\tilde d_2$ ($\tilde c_2$ and $\tilde s_2$),
or the left-handed sleptons, $\tilde e$ and $\tilde\nu_1$
($\tilde\mu$ and $\tilde\nu_2$), remain close to each other,  
a reflection of the sum rules Eqs. (\ref{indep1}) and (\ref{indep2}).

The spectrum of the U(1) model at the weak scale 
is shown in Fig. \ref{spec3}.
It is seen that the spectrum ends with relatively small $m_0$, since
for large enough $m_0$ the negative charges lead to a tachyonic spectrum.
The mass squared difference of squarks $\tilde d_L$ and $\tilde d_R$
(or $\tilde s_L$ and $\tilde s_R$) is seen to satisfy the 
sum rule Eq.~(\ref{u11}).

\section{Summary and discussion}

We have derived sum rules, and obtained the
sparticle mass spectra, in different AMSB models in which  the problem
of tachyonic slepton masses has been solved in qualitatively
different ways.  Interestingly enough, the sum rules obtained 
uncover many of the similarities and differences in these models.
On the one hand this helps us to tell whether a particular mass
spectrum is due to anomaly mediation, and on the other hand it allows us
to differentiate between various anomaly mediated models.

Apart from the well known feature of the AMSB gaugino sector,
the close mass degeneracy of the lightest chargino and neutralino, 
the sum rules that we have derived 
reveal another typical feature of the gaugino sector
in these models, namely the average mass difference between
charginos and neutralinos.
As shown in Fig. \ref{gauginos}
the average mass difference between the charginos and neutralinos is
typically negative in the AMSB type models.
This is contrary to the SUGRA type models, where the average mass
difference is positive.

Even if the mass spectrum of sparticles points towards the underlying
model being of AMSB type,
there could be a wide variety of different possibilities.
In such a situation, the sum rules obtained in this paper
would help in distinguishing between various models.
Comparison of the sfermion mass differences 
$M_{\tilde e_L}^2 - M_{\tilde e_R}^2$ and 
$M_{\tilde d_L}^2 - M_{\tilde d_R}^2$ can be crucial
in distinguishing  between the models considered in this paper. 
In the minimal AMSB model both these mass differences are small,
whereas in the gaugino assisted model slepton mass difference is
not small. In the AMSB model with extra U(1) and a light stop, 
the squark mass difference is considerable, but the slepton mass 
difference is small.
These conclusions from the tree-level sum rules of Section 2 have 
been verified by the one-loop  numerical results of Section 4.

In the minimal AMSB model,  $m_{H_u}$
has a focus point near the weak scale.  
We have shown that this attractive feature is, unfortunately,
not shared by more general AMSB models.

Finally, we should mention the possibility
that the introduction of $R$-parity violating couplings
can solve the tachyonic slepton mass problem. 
This is an interesting possibility, since if viable, one would have in
the supersymmetry breaking sector only those terms which arise
due to the anomaly mediation.
Although this idea is interesting, it turns out that it is not to easy 
to find suitable $R$-parity violating terms for its implementation.
In \cite{AD} a hierarchy between various $R$-parity
violating couplings was assumed, and it was
further assumed that these  couplings have quasi fixed points.

However, one of the couplings used contributes
to the neutrino masses \cite{dl}.
Thus it is subject to strict phenomenological bounds, 
due to which the scenario considered does not seem to be viable
in practice. On the other hand, if one assumes that the coupling 
involved in neutrino masses
is tiny, one may still generate large enough masses for all the
sleptons with the remaining two couplings used in~\cite{AD}, 
although the fixed point structure does not hold anymore.
Since one has in this case only the two unknown $R$-parity violating
couplings in addition to the mass parameter $m_{3/2}$, one can solve
for all these parameters and obtain different  sum rules. 
For the first two generations one has 14 masses, and thus
one gets 11 sum rules between the masses.
However, we think that this model is not very compelling
and do not list these sum rules here.\footnote{The couplings and the
mass parameter are given by 
\bea
&&\frac{66}{25}g_1^4\frac{m_{3/2}^2}{(16\pi^2)^2}= -(M_{\tilde u_R}^2-
M_{\tilde d_R}^2)+\sin^2\theta_W M_Z\cos 2\beta ,\nonumber\\
&&2\lambda_{231}\beta (\lambda_{231} )\frac{m_{3/2}^2}{(16\pi^2)}
=M_{\tilde e_R}^2-3
(M_{\tilde u_R}^2-M_{\tilde d_R}^2)+4\sin^2\theta_W M_Z\cos 2\beta ,
\nonumber\\
&&\lambda_{132}\beta (\lambda_{132} )\frac{m_{3/2}^2}{(16\pi^2)}=
M_{\tilde e_R}^2 +(M_{\tilde e_L}^2-M_{\tilde e_R}^2)-
(M_{\tilde u_R}^2-M_{\tilde d_R}^2)-(M_{\tilde d_L}^2-M_{\tilde
d_R}^2)
+\frac 23\sin^2\theta_W M_Z\cos 2\beta .\nonumber
\eea
}

\begin{flushleft} {\bf Acknowledgments} \end{flushleft}

\noindent
The authors thank the Academy of Finland
(project numbers 163394 and 48787) for financial support.
The work of P.N.P. is also
supported by the Department of Atomic Energy, India.
P.N.P. would like to thank the Alexander von Humboldt-Stiftung
and Professor H. Fraas for the hospitality extended to him at
Universit\"at W\"urzburg, where this work was completed.

\end{document}